\documentclass[pra,10pt,twocolumn,showpacs]{revtex4}
\usepackage[dvips]{epsfig}
\usepackage{float}
\usepackage{amsfonts}
\usepackage{amsmath}
\usepackage{amssymb}
\usepackage{enumerate}

\newcommand{\vecb}[1]{\mathbf{#1}}

\begin{document}

\title{An optical fiber-based probe for photonic crystal microcavities}

\author{Kartik Srinivasan}
\email{kartik@caltech.edu}
\thanks{\newline \indent This work has been submitted to the IEEE for possible publication.  Copyright may be transferred without notice, after which this version may be no longer accessible.}
\author{Paul E. Barclay}
\author{Matthew Borselli}
\author{Oskar Painter}
\affiliation{Department of Applied Physics, California Institute of Technology, Pasadena, California 91125.}
\date{\today}
\begin{abstract}
We review a novel method for characterizing both the spectral and spatial properties of resonant cavities within two-dimensional photonic crystals (PCs).  An optical fiber taper serves as an external waveguide probe whose micron-scale field is used to source and couple light from the cavity modes, which appear as resonant features in the taper's wavelength-dependent transmission spectrum when it is placed within the cavity's near field. Studying the linewidth and depth of these resonances as a function of the taper's position with respect to the resonator produces quantitative measurements of the quality factor ($Q$) and modal volume ($V_{\text{eff}}$) of the resonant cavity modes.  Polarization information about the cavity modes can be obtained by studying their depths of coupling when the cavity is probed along different axes by the taper. This fiber-based technique has been used to measure $Q \sim 40,000$ and $V_{\text{eff}} \sim 0.9$ cubic wavelengths in a graded square lattice PC microcavity fabricated in silicon. The speed and versatility of this fiber-based probe is highlighted, and a discussion of its applicability to other wavelength-scale resonant elements is given.  
\end{abstract}
\maketitle


\setcounter{page}{1}

\section{Introduction}
The quality factor ($Q$) and modal volume ($V_{\text{eff}}$) of an optical cavity quantify two of its most important properties, namely, the photon lifetime within the resonator and the per photon electric field strength, respectively.  $Q$ and $V_{\text{eff}}$ are central quantities in the study of a number of optical processes within microcavities \cite{ref:Chang}, such as coherent atom-photon coupling in cavity quantum electrodynamics (cQED) \cite{ref:Kimble2}, nonlinear processes such as stimulated Raman scattering \cite{ref:Spillane1}, and enhanced radiative processes in novel emitters \cite{ref:Purcell,ref:Gerard2,ref:Michler,ref:Solomon}.  Microcavities within two-dimensional photonic crystals \cite{ref:Painter3,ref:Ryu3,ref:Srinivasan4,ref:Akahane2} are distinguished by their ultra-small $V_{\text{eff}}$ values, approaching the theoretical limit of a cubic half-wavelength in the material ($(\lambda/2n)^3$).  When combined with recent improvements in fabrication and design that have led to experimental demonstrations of $Q$ factors well in excess of $10^4$ \cite{ref:Akahane2,ref:Srinivasan7,ref:Srinivasan8} and approaching $10^5$ \cite{ref:Noda_PECS}, these cavities become particularly interesting for a number of experiments, such as those involving strong coupling in cQED \cite{ref:Lev}, particularly in light of their inherent potential for chip-based integration for future experiments in cQED and quantum networks \cite{ref:Mabuchi}.       

Measurements of PC microcavities are not necessarily straightforward, however, in large part due to their micron-scale $V_{\text{eff}}$ values, which limit the ability to effectively couple to them from free-space or through prism-based techniques, as can be done for larger microresonators such as Fabry-Perot cavities \cite{ref:Kimble2} and microspheres \cite{ref:Collot}.  This difficulty has generally extended to other types of wavelength-scale cavities, such as small-diameter microdisks \cite{ref:Gayral,ref:Michler2}.  Typically, there have been two techniques to probe $Q$ factors in wavelength-scale cavities.  In the first, the microcavities are fabricated in an active emitter material (such as a quantum well or quantum dot epitaxy), the cavities are optically pumped, and the emitted resonance linewidth is studied sub-threshold, near material transparency \cite{ref:Gayral,ref:Srinivasan4,ref:Yoshie2}.  This technique is limited both by difficulties in accurately establishing the pump power at which transparency occurs and by the necessity that the cavity contain embedded emitters.  In particular, the latter requirement limits the variety of material systems in which the cavity can be fabricated (silicon, for example, would not be an option) and is not suitable for passive resonators in devices such as filters.  For such devices, a second technique, consisting of fabricating an on-chip in-plane waveguide to couple to the cavity, is often used \cite{ref:Little2,ref:Noda2}.  In this approach, the problem of coupling light into the cavity is shifted to that of coupling light into the on-chip waveguide, a technically less challenging problem that can be done through a number of end-fire based approaches.  A limitation of this technique is that it lacks a certain amount of flexibility due to the necessity of fabricating an in-plane waveguide for each cavity on the chip.  In addition, both this approach and the latter method described above do not provide a means to probe the $V_{\text{eff}}$ of the cavity.  To address this, several researchers have begun to investigate photonic crystal microresonators using near-field scanning optical microscopy (NSOM), taking advantage of the sub-wavelength resolution that can be achieved in such measurements to map the localization of the cavity modes\cite{ref:Okamoto,ref:Buchler}.
          
Here, we present an alternate technique \cite{ref:Srinivasan7}, which employs an $\it{external}$ waveguide to couple to the cavity, where the external waveguide is a tapered optical fiber.  Tapered optical fibers have been succesfully used in the past to excite the resonances of larger sized microcavities, such as silica microspheres \cite{ref:Knight,ref:Cai} and microtoroids \cite{ref:Armani}, and more recently, to excite the modes of a silicon-based PC waveguide \cite{ref:Barclay3,ref:Barclay4}.  In these implementations, phase-matching between the mode of the taper and the traveling wave mode of the resonator or the propagating mode of the waveguide was critical in achieving highly efficient coupling \cite{ref:Spillane2,ref:Barclay5}; in the former case, phase matching occurred primarily due to the silica-silica interface (same material index) between the taper and the microsphere, while in the latter, the dispersion of the PC waveguide was engineered to compensate for the disparate material indices ($n$=$3.4$ for silicon and $n=1.45$ for silica) and achieve a PC waveguide effective index that matched that of the taper. 

To study wavelength-scale cavities, we no longer rely on phase-matching, but rather just use the taper as a convenient means to produce a micron-scale evanescent field for sourcing and out-coupling the micron-scale cavity field.  Although such coupling might not be as efficient as phase-matched coupling, as we shall see, the power transfer is more than adequate enough to probe many of the important properties of the cavity.  By using an external waveguide as the coupling element, this method is inherently non-invasive, can be used to rapidly characterize multiple devices on a chip, and as we shall discuss below, the ability to vary the position of the taper with respect to the cavity (not an option for microfabricated on-chip waveguides) allows for quantitative investigation of not only the $Q$ factor but also $V_{\text{eff}}$.  Furthermore, the resonant coupling from the external waveguide is polarization selective, providing additional information about the cavity modes that is not easily obtainable through techniques such as NSOM.  Thus, in some respects, the versatility of the fiber-based approach described here makes this technique an optical analog to electrical probes used to study microelectronic devices.  

The body of this paper is partitioned into two main sections.  In Section \ref{sec:probe_setup}, we describe the basics of the probe setup, while in Section \ref{sec:PC_measurements}, measurements of fabricated PC microcavities using this optical fiber taper probe are presented.  In the concluding section (Section \ref{sec:conclusions}), we summarize the key points of the paper and discuss further applications of this method towards future studies of wavelength-scale cavities.        

\section{The optical fiber taper probe setup}
\label{sec:probe_setup}

\begin{figure}[ht]
\begin{center}
\epsfig{figure=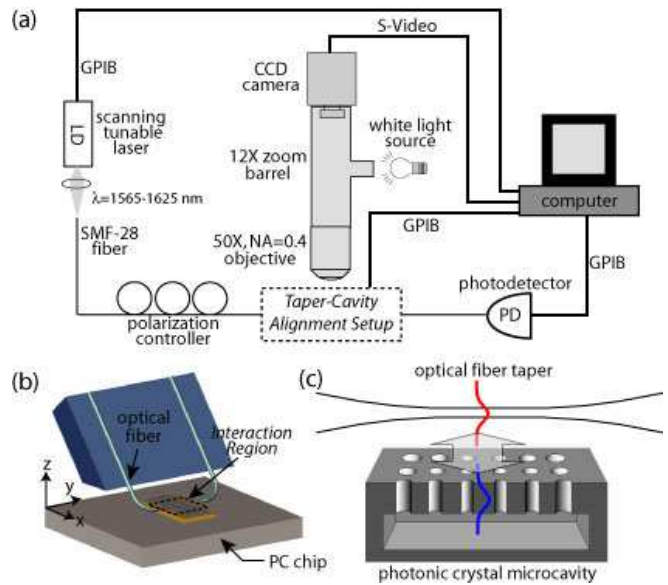, width=\linewidth}
\caption{(a) Experimental measurement setup for probing optical microcavities with fiber tapers. (b) Taper-cavity alignment setup (dashed boxed region in (a)).  The fiber taper is mounted in a u-shaped configuration and attached to a $\hat{z}$-axis stage with 50 nm resolution, while the underlying PC chip can be positioned in the $\hat{x}-\hat{y}$ plane with 50 nm resolution.  The fiber taper is spliced into the setup in (a), which allows for measurement of its wavelength-dependent transmission spectrum. (c) Zoomed-in depiction of the taper-cavity interaction region (dashed boxed region in (b)).}
\label{fig:taper_setup}
\end{center}
\end{figure}

The fiber tapers we use consist of a standard single mode optical fiber (9 $\mu$m core diameter, 125 $\mu$m cladding diameter) that has been simultaneously heated and stretched down to a minimum diameter ($d$) on the order of the wavelength of light ($\lambda$), so that for $\lambda\sim$1.6 $\mu$m as used in our experiments, $d\sim$1-2 $\mu$m.  To form the tapers used here, the heating mechanism is a hydrogen-based torch \cite{ref:Birks}, but other techniques such as use of a CO$_2$ laser have also been studied by other groups \cite{ref:Russell4}.  In a taper with a suitably adiabatic transition region, the insertion loss through the taper can be quite low; the tapers we typically fabricate have an insertion loss of $\sim 5-30$ $\%$.  The taper is mounted onto an acrylate block in a u-shaped configuration (Fig. \ref{fig:taper_setup}(b)), and the block is then fastened to a DC motor-controlled $\hat{z}$-axis stage with 50 nm step size resolution.  The microcavity chip (to be described in more detail below) is in turn mounted on a DC motor-controlled $\hat{x}-\hat{y}$-axis stage with 50 nm step size resolution; in this way, the fiber taper can be precisely aligned to a microcavity. The taper-cavity interaction region is imaged with a microscope onto a CCD camera.

The mounted taper is fusion spliced into the measurement setup (Fig. \ref{fig:taper_setup}(a)) so that a fiber-coupled scanning tunable laser with polarization-controlling paddle wheels is connected to its input and an InGaAs photodetector measures its output.  The laser and photodetector output are attached to a computer via GPIB interfaces, so that the wavelength-dependent transmission of the taper can be recorded.  In addition, the motorized stages on which the taper and PC chip are mounted are also GPIB-controlled, so that the taper transmission spectrum can be monitored as a function of the taper's position with respect to the cavity.  When the taper is laterally aligned over the central region of the cavity and positioned vertically within the cavity's near-field (typically $<$ 1 $\mu$m), the cavity modes appear as resonances within this transmission spectrum.  As we shall discuss in Section \ref{sec:PC_measurements}, measurements of the linewidth and depth of these resonances as a function of the taper's position with respect to the cavity can give us quantitative estimates of the cavity's $Q$ and $V_{\text{eff}}$.  

\begin{figure}[ht]
\begin{center}
\epsfig{figure=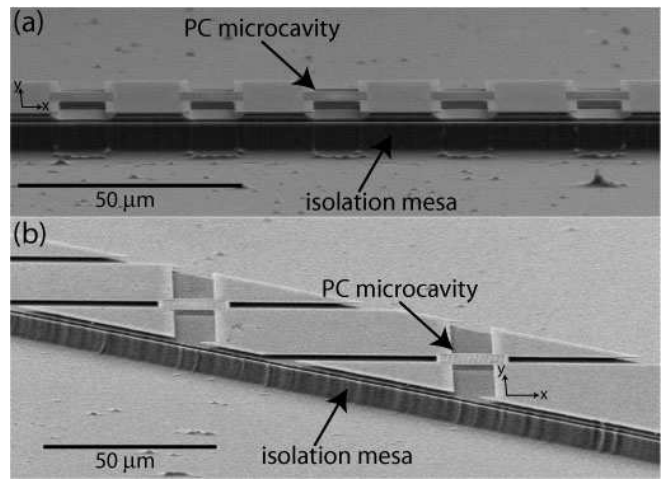, width=\linewidth}
\caption{(a)-(b) Scanning electron microscope (SEM) images of photonic crystal microcavity arrays fabricated in silicon-on-insulator.  The undercut PC microcavities are fabricated in a linear array that is isolated from the rest of the chip by several microns.  The devices in (a) have additional material removed along the $\hat{y}$-axis of the cavity, to allow for the fiber optic taper to be aligned along that axis.  The devices in (b) have additional material removed along both the $\hat{x}$ and $\hat{y}$ axes, to allow probing along either axis.}   
\label{fig:isolated_samples}
\end{center}
\end{figure}

In principle, the curvature of the looped taper shown in Fig. \ref{fig:taper_setup}(b) could be made large enough so that only a small region of the fiber interacts with the PC chip, and the taper could then be used to probe a full two-dimensional (2D) array of cavities on a chip. For the fiber tapers we have used, this region is typically around $10$ mm in length, and is roughly equal to the length of the tapered region of the fiber as defined when the taper is formed.  Mounting the taper in this fashion naturally keeps it under tension and prevents the taper position from excessively fluctuating due to environmental factors (such as fluctuating air currents in the laboratory).  Using more tightly looped tapers requires either a reduction in the length of taper during its fabrication (while still maintaing enough adiabaticity to keep losses low), or some level of isolation from these environmental factors.  One result of this relatively long $10$ mm interaction length is that testing of a 2D array is not feasible, and linear (1D) arrays of devices are tested.  In addition, because coupling to the cavity requires the taper to be positioned within a micron of the center region of cavity, control of the tip and tilt of the sample with respect to the taper is necessary; this is accomplished through use of a goniometer stage mounted to the motor-controlled $\hat{x}-\hat{y}$ sample stage.  Finally, to prevent the taper from interacting with extraneous portions of the chip, the cavities are isolated to a mesa stripe that is several microns above the rest of the sample surface (Fig. \ref{fig:isolated_samples}); this step is described in more detail in the following section.

\section{Measurements of PC microcavities using optical fiber tapers}
\label{sec:PC_measurements}

\begin{figure}[ht]
\begin{center}
\epsfig{figure=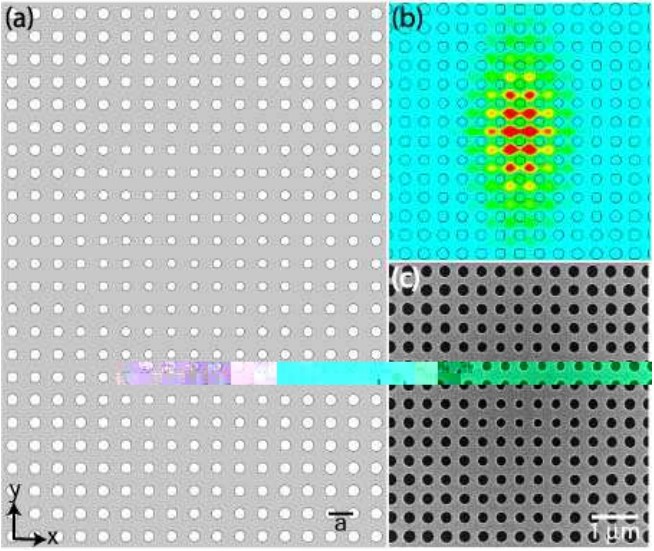, width=\linewidth}
\caption{(a) Graded square lattice photonic crystal cavity designed in Ref. \cite{ref:Srinivasan1}. (b) Finite-difference time-domain (FDTD) calculated mangetic field amplitude in the center of the slab for the $A^{0}_{2}$ mode of interest (c) SEM image of a fabricated device.}   
\label{fig:design_fab_cavity_top}
\end{center}
\end{figure}

The particular PC microcavities we use in these experiments follow the graded square lattice design (see Fig. \ref{fig:design_fab_cavity_top}) developed in Ref. \cite{ref:Srinivasan1}, and are predicted to sustain a mode (labeled $A^{0}_{2}$, due to its fundamental nature and symmetry properties) with a finite-difference time-domain (FDTD) calculated $Q$ factor on the order of $10^5$ with $V_{\text{eff}}\sim (\lambda/n)^3$.  To fabricate these undercut devices in silicon-on-insulator wafers (SOI) while incorporating the isolation mesas mentioned above, the following processing steps are performed: (1) electron beam lithography of the PC pattern and accompanying cutouts for removal of additional material from the mesa, (2) SF$_6$/C$_4$F$_8$-based inductively-coupled plasma reactive ion etching (ICP-RIE) through the silicon membrane layer, (3) removal of the electron beam resist, (4) photolithography to define a mesa stripe that intersects the electron beam defined cutouts, (5) removal of material surrounding the mesa (dry etching of the top silicon, underlying oxide, and substrate silicon layers), (6) removal of the photoresist, and (7) wet etch (hydrofluoric acid) of the underlying oxide layer to form a free-standing membrane.  Fig. \ref{fig:isolated_samples} shows scanning electron microscope (SEM) images of fully-processed devices; in addition to being isolated to the mesa stripe, additional cutout material (defined in step (1) above) surrounding each cavity has been removed to ensure that the taper interacts only with the cavity.  The thickness of the membrane layer in our SOI samples is 340 nm, and prior to undercut, the underlying oxide thickness is 2 $\mu$m.  Fig. \ref{fig:design_fab_cavity_top} shows a top view of the nominal cavity design (Fig. \ref{fig:design_fab_cavity_top}(a)) and a fabricated device (Fig. \ref{fig:design_fab_cavity_top}(c)).  Within the linear array of devices, we fabricate two or three different lattice constants (on the order of $a$=$400$ nm, so that $a/\lambda \sim 0.25$ for $\lambda$=1600 nm, to be consistent with simulation results from Ref. \cite{ref:Srinivasan1}), and for a given $a$, the average hole radius ($\bar{r}/a$) is varied.  The combination of varying $a$ and $\bar{r}/a$ allows us to easily tune the cavity resonances through the range of our scanning tunable laser, $\lambda$=1565-1625 nm.              

The coupling between the tapered fiber waveguide and the PC microcavity can be understood using, for example, the coupling of modes in time approach as in Manolatou, et al. \cite{ref:Manolatou}.  The degree to which the taper mode couples to the cavity mode is a function of the parameter $\kappa$, which is approximately given by the field overlap between the two modes over their interaction length.  More explicitly, for the taper aligned along the $\hat{y}$-axis of the cavity, it is given by: 

\begin{equation}
\label{eq:kappa}
\begin{split}
\kappa=-\frac{i\omega{\epsilon}_0}{4}\int_{0}^{L}e^{-i{\beta}y}dy\biggl(\iint\limits_{A_c}(n^2-n_{c}^2){\vecb{E}}_{c}^{\ast}\cdot{\vecb{E}}_{t}dxdz + \\ \iint\limits_{A_t}(n^2-n_{t}^2){\vecb{E}}_{c}^{\ast}\cdot{\vecb{E}}_{t}dxdz{\biggr)},
\end{split}
\end{equation}   

\noindent where $\omega$ is the resonant cavity mode frequency, $\epsilon_{0}$ is the permitivitty of free space, $\beta$ is the propagation constant of the taper mode, $n$ is the refractive index of the background air, $n_{c}$ ($n_{t}$) is the refractive index profile of the cavity (taper), $\vecb{E}_{c}$ ($\vecb{E}_{t}$) is the electric field vector of the cavity (taper), $L$ is the interaction length of the coupling, and the integrals over $A_{c}$ and $A_{t}$ are two-dimensional integrals over the $x$-$z$ cross-section within the cavity and taper, respectively, for a given $y$ value within the interaction length.  From this formula, we see that $\kappa$ is dependent upon: (i) the magnitude of the overlap between $\vecb{E}_{c}$ and $\vecb{E}_{t}$, (ii) the relative phase between $\vecb{E}_{c}$ and $\vecb{E}_{t}$, and (iii) the degree to which $\vecb{E}_{c}$ and $\vecb{E}_{t}$ share a common direction of polarization.  

\begin{figure}[ht]
\begin{center}
\epsfig{figure=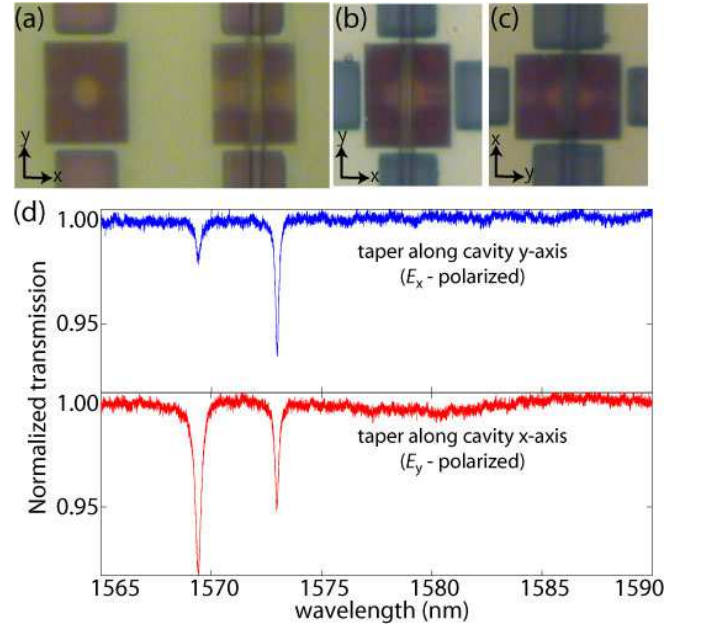, width=\linewidth}
\caption{(a) Optical micrograph image of a fiber taper aligned along the $\hat{y}$-axis of one of the cavities from Fig. \ref{fig:isolated_samples}(a).  (b)-(c) Optical micrograph images of a fiber taper aligned along the $\hat{y}$ and $\hat{x}$ axes, respectively, of one of the cavities from Fig. \ref{fig:isolated_samples}(b). (d) Normalized taper transmission when the taper is $\sim 350$ nm above a PC cavity of the type shown in Fig. \ref{fig:isolated_samples}(b); (top) Taper aligned along the cavity's $\hat{y}$-axis (bottom) Taper aligned along the cavity's $\hat{x}$ axis.}   
\label{fig:taper_over_cavity}
\end{center}
\end{figure}

Considering the latter point, for the taper aligned along the $\hat{y}$-axis of the cavity, as in Fig. \ref{fig:taper_over_cavity}(a), the polarization-controlling paddle wheels are used to select a linearly polarized state of the taper whose dominant field component is aligned along the $\hat{x}$-axis of the cavity.  Thus, modes which couple most strongly will have their $\hat{x}$ field component overlap strongly with the $\hat{x}$ component of the taper field.  By aligning the taper along a different axis of the cavity, polarization selectivity can be realized.  For the devices in Figs. \ref{fig:isolated_samples}(a) and \ref{fig:taper_over_cavity}(a), this is not possible, as aligning the taper along the $\hat{x}$ axis will couple the taper to multiple cavities.  However, by fabricating the cavities at a 45$^{\circ}$ angle with respect to the isolation mesa stripe, as in Fig. \ref{fig:isolated_samples}(b), the taper can be aligned along either of the orthogonal cavity axes without coupling to multiple devices.  In Fig. \ref{fig:taper_over_cavity}(b)-(c), we show the optical fiber taper aligned parallel to the $\hat{y}$ and $\hat{x}$ axes of one of the cavities from Fig. \ref{fig:isolated_samples}(b).  When the taper is laterally aligned to the center of the cavity and brought vertically close to it, we observe the cavity's resonances.  As shown in Fig. \ref{fig:taper_over_cavity}(d), the coupling is polarization selective, so that those resonances with dominant cavity field component $E_{x}$ couple most strongly when the taper is aligned along the $\hat{y}$ axis of the cavity (Fig. \ref{fig:taper_over_cavity}(b)), while those with dominant cavity component $E_{y}$ couple most strongly when the taper is aligned along the cavity's $\hat{x}$ axis (Fig. \ref{fig:taper_over_cavity}(c)).  Thus, the shorter wavelength resonance in Fig. \ref{fig:taper_over_cavity}(d) is more strongly polarized along the $\hat{y}$-axis, while the longer wavelength resonance is more strongly polarized along the $\hat{x}$-axis.  The coupling depths of a few percent are typical values, and were found to be adequate to achieve a sufficient signal-to-noise ratio for all of the measurements discussed in this work.

For future microcavity-enhanced experiments, the $A^{0}_{2}$ mode of this graded square lattice microcavity is of particular interest, on account of its predicted ultra-small $V_{\text{eff}}$ and high $Q$ factor.  To experimentally locate a device for which this mode appears within the scan range of the laser we use (1565-1625 nm), we rely on the fact that it is the fundamental (and hence lowest frequency) mode within the region of $\omega-k$ space under consideration.  In particular, for a given lattice constant $a$, we begin testing by examining the device with the largest $\bar{r}/a$.  Typically, the resonances of this device are at frequencies that are above that covered by the laser scan range, and hence, no resonances are observed in the taper's transmission spectrum.  We then move on and test the next device in the array, which has a slightly smaller $\bar{r}/a$ and is thus predicted to have lower frequency resonances than the previous device.  This process is continued until we find a device for which a resonance is seen in the transmission spectrum.  The first resonance which appears in the transmission spectrum (measuring from lowest frequency to highest frequency) is the $A^{0}_{2}$ mode; this identification procedure relies on having only small changes in $\bar{r}/a$ between successive devices in the array, so that the cavity resonances can be smoothly tuned from frequencies above the laser scan range to frequencies within the scan range.  However, this identification of the $A^{0}_{2}$ mode can be confirmed, both by comparing the measured resonance frequency with that predicted from FDTD simulations using the SEM-measured hole sizes, and as discussed later, by comparing the spatial localization of the cavity mode with that predicted from simulations. 

\begin{figure}[ht]
\begin{center}
\epsfig{figure=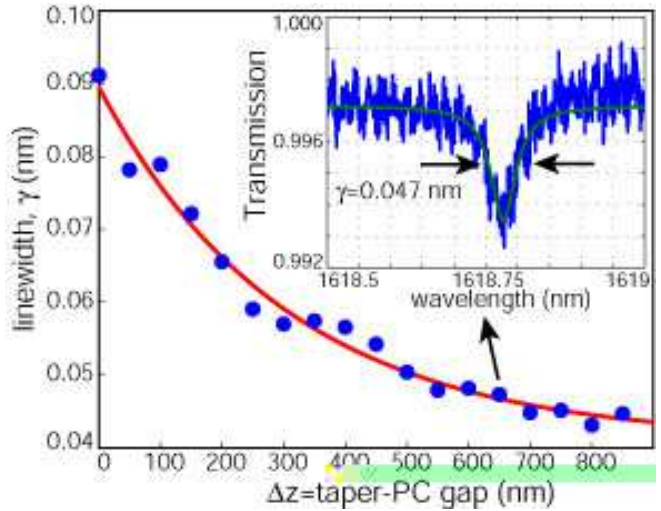, width=\linewidth}
\caption{Measured linewidth (blue dots) versus taper-cavity gap (${\Delta}z$) for the $A^{0}_{2}$ mode ($a/\lambda_{0} \sim 0.263$) in a sample with $a$=425 nm.  The solid red curve is a fit to the experimental data. (inset) Normalized taper transmission versus wavelength when the taper is $650$ nm above the cavity surface. Figure taken from Ref. \cite{ref:Srinivasan7}.}
\label{fig:z_scan_data}
\end{center}
\end{figure}

Having identified a device for which the $A^{0}_{2}$ mode appears within the scan range of the laser, we next examine its $Q$ factor and spatial localization (related to $V_{\text{eff}}$).  To demonstrate the techniques used, we rely upon data orignally presented in Ref. \cite{ref:Srinivasan7}.  In the inset of Fig. \ref{fig:z_scan_data}, we show a wavelength scan of the taper transmission showing the resonance dip of the $A^{0}_{2}$ mode for a device within the array shown in Fig. \ref{fig:isolated_samples}(b), where the vertical taper-cavity separation $\Delta z$=650 nm. An initial estimate of the $Q$ of this resonance (centered at wavelength $\lambda_0$) is given by measuring its linewidth $\gamma$, with $Q=\lambda_0/\gamma$.  For this device, $\lambda_{0}\sim 1618.75$ nm and $\gamma\sim$ 0.047 nm (where these values are determined by fitting the data to a Lorentzian curve), giving an estimate of $Q\sim 34,400$.  This $Q$ is a lower bound for the cold-cavity $Q$, due to the taper's loading effects on the cavity, which cause its linewidth to broaden.  Parasitic taper loading effects could include coupling to radiation modes or higher-order taper modes.  These loading effects are expected to diminish as $\Delta z$ is increased and the taper is positioned further and further above the cavity, until a regime is reached where the taper does not significantly affect the cavity mode and $\gamma$ does not change with increasing $\Delta z$.  This asymptotic linewidth $\gamma_{0}$ gives us our best estimate for the cold-cavity $Q$\footnote{Note that some parasitic taper loading effects may not diminish as a function of $\Delta z$ as rapidly as does the coupling between the fundamental taper mode and cavity mode of interest. This could prevent the measured linewidth $\gamma$ from reaching an asymptotic value as a function of $\Delta z$.  In such cases, the best estimate of $\gamma_0$ is the linewidth for as large a $\Delta z$ as can be reliably measured.}.   For the current device under consideration, the variation of $\gamma$ with $\Delta z$ is shown in Fig. \ref{fig:z_scan_data}.  If this curve is fit to a simple monoexponential function, a cold-cavity linewidth $\gamma_{0}$=$0.041$ nm is obtained, which is essentially identical to the directly measured linewidth when ${\Delta}z \gtrsim 800$ nm, and corresponds to a cold-cavity $Q \sim 39,500$.  This measured $Q$ and normalized frequency of $a/\lambda_{0}\sim0.263$ correspond relatively well with the FDTD-predicted $Q\sim56,000$ and $a/\lambda_{0}\sim0.266$ for a device with hole radii similar to that measured by the SEM for this device.    

As mentioned earlier, the extremely small volumes to which light is confined within PC microcavities is one of their distinguishing advantages over other optical microcavities, and is of critical importance in many applications, as the per photon electric field strength within the cavity is proportional to $1/\sqrt{V_{\text{eff}}}$.  The ability to experimentally confirm such tight spatial localization using the same probe that maps the spectral properties (such as the $Q$) of the cavity modes is an important demonstration of the versatility of the optical fiber taper.  Here, the same near-field probe is used to both excite the PC cavity modes and to map their spatial profile.  Other works employing evanescent coupling from eroded monomode fibers to excite silica microsphere whispering-gallery modes have used a secondary fiber tip to collect and map the mode profiles \cite{ref:Knight2}.

\begin{figure}[ht]
\begin{center}
\epsfig{figure=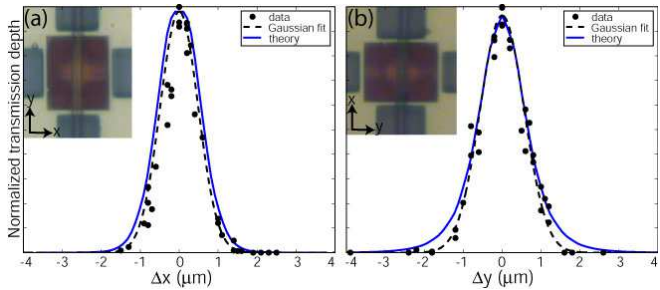, width=\linewidth}
\caption{Mode localization data for the cavity whose $Q$ was measured in Fig. \ref{fig:z_scan_data}. The measured normalized taper transmission depth (black dots) is plotted as a function of taper displacement along the (a) $\hat{x}$-axis and (b) $\hat{y}$-axis of the cavity. The dashed line in (a)-(b) is a Gaussian fit to the data while the solid line is a numerically calculated coupling curve based upon the FDTD-generated cavity field and analytically determined fundamental fiber taper mode (taper diameter $d\sim 1.7$ $\mu$m).  Figure taken from Ref. \cite{ref:Srinivasan7}.}
\label{fig:lateral_scan_data}
\end{center}
\end{figure}

The spatial localization of the cavity mode is easily probed by examining the depth of coupling between it and the taper as the taper is laterally scanned above the surface of the cavity.  In the insets of Fig. \ref{fig:lateral_scan_data}(a)-(b), we show the fiber taper aligned along the $\hat{y}$ and $\hat{x}$ axis of the photonic crystal microcavity whose $Q$ was measured to be $\sim 40,000$.  The position of the taper is varied along the $\hat{x}$ and $\hat{y}$ axes of the cavity, respectively, allowing for measurements of the depth of coupling along these two orthogonal cavity directions.  The depth of the resonant transmission dip for the $A^{0}_{2}$ cavity mode versus taper displacement is shown in Figs. \ref{fig:lateral_scan_data}(a) and \ref{fig:lateral_scan_data}(b), respectively.  These measurements show the mode to be well-localized to a micron-scale central region of the cavity, giving experimental confirmation that the $A^{0}_{2}$ mode of this cavity is both high-$Q$ and small $V_{\text{eff}}$.  As might be expected, they do not reveal the highly oscillatory cavity near-field, but instead an envelope of the field, due to the relatively broad taper field profile.  To compare this experimental data with FDTD calculations, we consider a simple picture of waveguide-cavity coupling \cite{ref:Manolatou}, where the coupling coefficient ($\kappa$) is approximated, to save computation time, by taking the field overlap of the phase-matched Fourier components of the FDTD-generated cavity field with the analytically-determined taper field.  The calculated resonant transmission depth as a function of taper displacement is shown in Figs. \ref{fig:lateral_scan_data}(a)-(b) as a solid line and agrees closely with the measured data, providing further confirmation that the mode studied is indeed the $A^{0}_{2}$ mode of interest.  Assuming that the cavity mode is localized to the slab in the $\hat{z}$-direction (a good assumption based upon measurements that show the depth of coupling between the taper and cavity mode decreases exponentially as the taper-cavity separation is increased), the close correspondence between the measured and calculated in-plane localization indicates that $V_{\text{eff}} \sim 0.9 (\lambda_{c}/n)^3$ for this high-$Q$ mode, where this $V_{\text{eff}}$ value was calculated through FDTD simulations which take into account the SEM-measured hole radii for this device.    

\begin{figure}[ht]
\begin{center}
\epsfig{figure=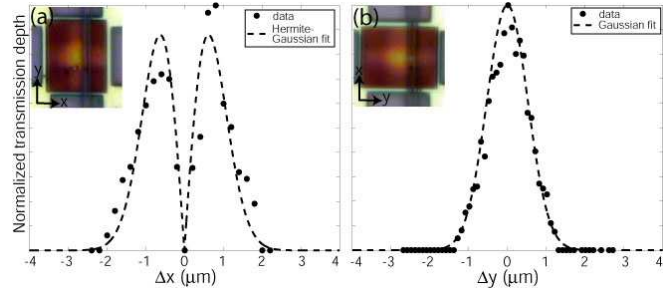, width=\linewidth}
\caption{Mode localization data for a higher order mode of the graded square lattice PC cavity. The measured normalized taper transmission depth (black dots) is plotted as a function of taper displacement along the (a) $\hat{x}$-axis and (b) $\hat{y}$-axis of the cavity. The dashed line in (a) is a Hermite-Gaussian fit to the data and the dashed line in (b) is a Gaussian fit to the data.}
\label{fig:lateral_scan_data_higher_order}
\end{center}
\end{figure}

Similar measurements of the higher-frequency resonant modes of the PC microcavity indicate that they are more delocalized in-plane in comparison to the $A^{0}_{2}$ mode, and sometimes contain multiple lobes within their coupling curves, as one might expect for higher-order modes of the cavity.  As an example of this, we show in Fig. \ref{fig:lateral_scan_data_higher_order}(a)-(b) the depth of coupling to a higher order mode as a function of the taper's position along the $\hat{x}$ and $\hat{y}$ axis of the cavity, respectively.  The node that appears within the coupling curve in Fig. \ref{fig:lateral_scan_data_higher_order}(a) results from the cavity and taper modes being precisely phased so that the integral determining $\kappa$ in Equation \ref{eq:kappa} is zero, and is a result of the measurement being sensitive to the fields within the cavity and taper rather than their intensities.  

\begin{figure}[ht]
\begin{center}
\epsfig{figure=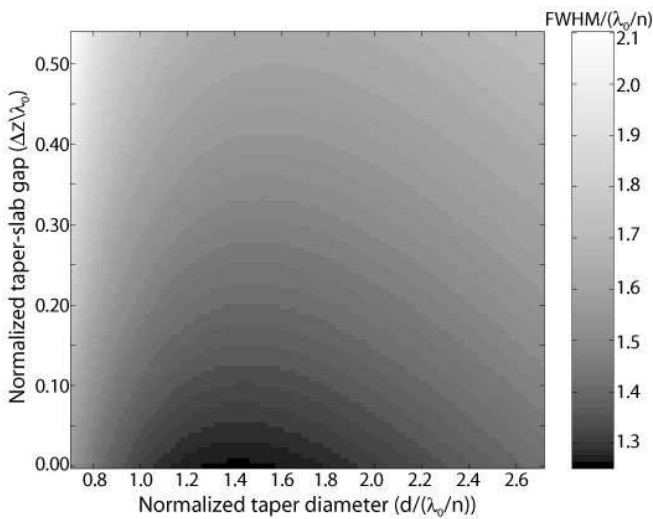, width=\linewidth}
\caption{Resolution of the fiber taper.  Plot of the normalized full-width at half-maximum (FWHM/($\lambda_0/n)$) of the dominant taper electric field component at the center of the PC slab, as a function of normalized taper-slab gap (${\Delta}z/\lambda_0$) and taper diameter ($d/(\lambda_0/n)$). $\lambda_0$ is the operating wavelength of the taper (and the resonant frequency of the cavity mode), and $n\sim1.45$ is its material refractive index.}
\label{fig:taper_field_size}
\end{center}
\end{figure}

With the exception of such cases where there is phase cancellation in $\kappa$, the resolution of the fiber taper probe is limited by the transverse profile of the taper mode.  This is the reason why the measured coupling curves give an envelope of the cavity field rather than displaying its oscillatory nature; in the measurements of Fig. \ref{fig:lateral_scan_data}, for example, the calculated full-width at half-maximum (FWHM) of the dominant taper field component at the center of the PC slab is $\sim\lambda_0$, while the cavity mode oscillates on the scale of a lattice constant $a$ (Fig. \ref{fig:design_fab_cavity_top}(b)), and $a/\lambda_{0}\sim0.25$.  The taper used in these measurements had a diameter $d\sim1.7$ $\mu$m ($d/(\lambda_0/n)\sim1.52$, where $n\sim1.45$ is the refractive index of the silica taper), and intuitively, it might be expected that better resolution could be achieved by further reducing its diameter.  However, for the relatively small taper diameters ($\sim \lambda_{0}$) with which we operate, we note that the waveguiding properties of the taper begin to degrade below some minimum diameter so that, even if a smaller taper is used, it does not necessarily confine the mode any more tightly than a larger taper would.  To better illustrate this, in Fig. \ref{fig:taper_field_size}, we plot the calculated normalized FWHM of the dominant taper field component at the center of the PC slab for varying normalized taper diameter ($d/(\lambda_0/n)$) and taper-PC slab separation (${\Delta}z/\lambda_0$).  As expected, the smallest FWHM$\sim 1.23(\lambda_0/n)$ occurs when ${\Delta}z/\lambda_0$=0, that is, when the taper is touching the slab\footnote{In practice, a non-zero ${\Delta}z$, on the order of $\sim 250$ nm for $\lambda_0$=1.6 $\mu$m, is preferable for doing spatial localization measurements.  This is due both to the relatively large amount of insertion loss that occurs when the taper touches the cavity, and also to allow the taper to be freely moved above the cavity.}.  We also see that reducing $d/(\lambda_0/n)$ below some minimum value begins to broaden the FWHM.  Thus, for the spatial localization measurements, using smaller tapers will not appreciably improve the resolution of the measurement.  Possibilities for future improvement might consist of partially aperturing the taper field (perhaps through a metallic coating on the taper), or forming the waveguide probe from a higher index material.   

\section{Discussion and Conclusions}
\label{sec:conclusions}

In the preceding paragraphs, we described how an optical fiber taper with a minimum diameter on the order of a wavelength can be used as a probe of ultra-small, wavelength-scale microcavities, where the particular cavities we considered were the PC microcavities originally studied in Ref. \cite{ref:Srinivasan7}.  The fiber taper can be used to probe both the spectral and spatial properties of the cavity's resonant modes, allowing for investigation of both modal quality factors and spatial localizations, giving quantitative estimates of $Q$ $\emph{and}$ $V_{\text{eff}}$.  As many microcavity-enhanced processes scale as a function of both of these parameters, being able to experimentally investigate each of them is an important attribute of this probing technique.  In addition, by aligning the taper along different directions above the cavity, polarization-sensitive information may be obtained.  Knowledge of a mode's spectral position, polarization, $Q$, and $V_{\text{eff}}$ will in many cases be enough to unambiguously determine the identity of the mode in comparison to simulation or theoretical results.  

Another important aspect of this technique is the speed with which measurements can be made.  In particular, the critical alignment step required in this work is making sure that the taper is not angled with respect to the surface of the chip, to ensure that coupling only occurs between the taper and the cavity, and not some extraneous part of the chip.  Once this is done, and once the taper is aligned along the desired axis of the cavity, all of the devices within an array can be rapidly tested, and the spectral positions of resonances in successive devices can be determined within tens of seconds or faster.  As an illustration of this, a movie showing the testing of two adjacent PC cavities has been made and is freely available on the internet (http://copilot.caltech.edu/research/PC\_cav.avi).  This ability to easily probe an array of devices on a chip greatly speeds up the testing process and shortens the turnaround time between device fabrication and measurement.  Furthermore the simplicity of the measurement technique is another attractive feature; a single fiber taper serves as both the excitation and collection probe, and the taper is physically robust enough (will not break) so that no active servo control of the taper position is required to prevent it from touching the sample surface (in contrast to the more delicate probes used in NSOM techniques \cite{ref:Buchler}).   

Finally, we note that the optical fiber taper probe can be used to examine the spectral and spatial properties of a number of wavelength-scale microcavities, and is not limited to just PC microcavities.  The suitability of the fiber taper as a probe for a given microcavity will in large part be determined by the overlap between the cavity and taper fields; simply put, if that overlap is sufficiently large, an appreciable amount of power can be transferred from the taper to the cavity even without phase-matching (in general, phase-matching will not be achieved, because of the index mismatch between the silica fiber taper and the high refractive index semiconductors typically used in wavelength-scale cavities).  For the $A^{0}_{2}$ PC microcavity mode studied in this work, the depth of coupling is typically limited to $\sim$ 10-20 $\%$, and at maximum levels of coupling, the cavity $Q$ is degraded due to the taper loading effects seen in Fig. \ref{fig:z_scan_data}.  However, due to the low-loss nature of the optical fiber tapers (insertion losses are routinely as low as 10 $\%$), this is still a significant amount of coupling into the cavity, and from the measurement standpoint, coupling levels of a few percent are easily adequate to discern cavity resonances in the taper's transmission, and to then probe the $Q$ and $V_{\text{eff}}$ of the cavity.  For applications requiring highly efficient power transfer from the fiber into the cavity, other approaches using an intermediate photonic crystal waveguide coupler have been developed \cite{ref:Barclay6}. 

As an example of the application of this probing technique to other types of wavelength-scale microcavities, fiber tapers have recently been used \cite{ref:Borselli} to probe the spectral and spatial properties of whispering gallery modes in $d<10$ $\mu$m diameter silicon microdisks.  Due to their large radiation-limited $Q$ factors ($>10^{15}$), measurement of the $Q$ in fabricated microdisks is a simple and elegant way to determine etch-induced and bulk material losses within a given materials system, allowing one to optimize an etching process to create the lowest loss devices.  Because the fiber taper measurement is a passive measurement (light-emitting material is not required), this probing technique provides optical access to materials systems, such as silicon, which otherwise could only be accessed via end-fire coupling to microfabricated on-chip waveguides.   

\section{Acknowledgements}
\label{Acknowledgements}
This work was partly supported by the Charles Lee Powell Foundation. K.S. thanks the Hertz Foundation and M.B. thanks the Moore Foundation, NPSC, and HRL Laboratories for their graduate fellowship support.

\end{document}